\documentclass[aip, apl, reprint, floatfix]{revtex4-1}

\usepackage{graphicx}
\usepackage{amsmath}

\usepackage[utf8]{inputenc}
\usepackage[T1]{fontenc}

\DeclareUnicodeCharacter{03BC}{µ}  
\DeclareUnicodeCharacter{2002}{ }  
\DeclareUnicodeCharacter{2009}{ }  
\DeclareUnicodeCharacter{2005}{ }  
\DeclareUnicodeCharacter{2212}{-}  
\DeclareUnicodeCharacter{2A7D}{<=}
\DeclareUnicodeCharacter{03C9}{w}

\makeatother
\usepackage{upgreek}

\begin{document}
\frenchspacing


\author{Annabelle Bricout}
\email{annabelle.bricout@universite-paris-saclay.fr}
\affiliation{Centre de Nanosciences et de Nanotechnologies, Université Paris-Saclay, CNRS, Palaiseau, 91120, France}
\author{Mathieu Bertrand}
\author{Philipp Täschler}
\author{Barbara Schneider}
\affiliation{Institute for Quantum Electronics, ETH Zurich, 8093 Zürich, Switzerland}
\author{Victor Turpaud}
\affiliation{Centre de Nanosciences et de Nanotechnologies, Université Paris-Saclay, CNRS, Palaiseau, 91120, France}
\author{Stefano Calcaterra}
\author{Davide Impelluso}
\author{Marco Faverzani}
\affiliation{L-NESS, Politecnico di Milano, Dipartimento di Fisica, Polo di Como, via Anzani 42, 22100 Como, Italy}
\author{David Bouville}
\author{Jean-René Coudevylle}
\author{Samson Edmond}
\author{Etienne Herth}
\author{Carlos Alonso-Ramos}
\author{Laurent Vivien}
\affiliation{Centre de Nanosciences et de Nanotechnologies, Université Paris-Saclay, CNRS, Palaiseau, 91120, France}
\author{Jacopo Frigerio}
\author{Giovanni Isella}
\affiliation{L-NESS, Politecnico di Milano, Dipartimento di Fisica, Polo di Como, via Anzani 42, 22100 Como, Italy}
\author{Jérôme Faist}
\affiliation{Institute for Quantum Electronics, ETH Zurich, 8093 Zürich, Switzerland}
\author{Delphine Marris-Morini}
\affiliation{Centre de Nanosciences et de Nanotechnologies, Université Paris-Saclay, CNRS, Palaiseau, 91120, France}

\title{On-chip pulse generation at 8 $\boldsymbol{\upmu}\text{m}$ wavelength}

\date{\today}

\begin{abstract}
The mid-infrared spectral region holds growing importance for applications such as gas sensing and spectroscopy. Although compact ultrashort pulse laser sources are essential to enable these applications, their realization in this spectral range remains an open challenge. We demonstrate an integrated approach to generate pulses in the mid-infrared based on chirped Bragg gratings engineered to compensate for the group delay dispersion of quantum cascade laser frequency comb sources. SiGe graded-index photonic circuits are used for operation around 8 $\upmu\text{m}$ wavelength. With this approach, pulses as short as 1.39 picoseconds were obtained, marking a key step towards fully integrated ultrashort pulse sources in the mid-infrared. 
\end{abstract}

\maketitle 
\section{Introduction}
 Up to now, ultrashort high power pulse generation in the mid-infrared wavelength range (2-20 $\upmu\text{m}$) relied mainly on bulky and expensive techniques\cite{pires_ultrashort_2015,sugiharto_generation_2008} based on difference-frequency generation\cite{murray_highly_2016,kaindl_broadband_1999}, optical parametric oscillators\cite{fraser_generation_1997,kumar_high-power_2015} or optical parametric amplification\cite{chalus_mid-ir_2009,liang_high-energy_2017}. However, photonic integration in the mid-infrared has recently gained significant interest, as this wavelength range corresponds to the "fingerprint region", enabling numerous applications such as infrared spectroscopy\cite{haas_advances_2016,faghihzadeh_fourier_2016} and gas sensing\cite{tombez_methane_2017,dean_methane_2018}. Compact ultrashort laser sources in the mid-infrared are of high interest as they would enable compact broadband solutions for many applications such as strong-field physics\cite{woodbury_laser_2018}, multi-species chemical detection\cite{abbas_fourier_2021} and standoff sensing\cite{kilgus_mid-infrared_2018}.

Quantum cascade lasers (QCL) are compact and flexible mid-infrared light sources that can operate from the mid-infrared up to the terahertz wavelength range\cite{faist_quantum_1994,vitiello_quantum_2015}. They can exhibit frequency comb behaviors\cite{hugi_mid-infrared_2012} with a temporal intensity profile that does not consist of short pulses as a mode-locked laser but an almost continuous profile instead\cite{villares_quantum_2015} with watt-level of power. In fact, QCLs exhibit very fast gain dynamics making it very difficult to produce pulses by passive mode-locking\cite{villares_quantum_2015} and hard to achieve high powers. This raises the necessity to compress the pulse externally instead. Interestingly, QCL frequency modulated combs exhibit a quadratic phase profile\cite{singleton_evidence_2018} making them excellent candidates for pulse compression\cite{saleh_fundamentals_2013}, and sub-picosecond pulses were obtained using a free-space external phase compensation stage\cite{singleton_pulses_2019,taschler_femtosecond_2021}. Considering the foreseen impact of mid-infrared photonics in sensing and spectroscopy, there is currently a huge interest to develop on-chip pulse compression of QCL frequency combs.

Several materials are under investigation for mid-infrared integration. Among them are III-V semiconductors\cite{dely_unipolar_2024,zhang_mid-infrared_2022} or chalcogenides\cite{yu_broadband_2014}. However, a silicon (Si) based platform is very interesting to benefit from complementary metal oxyde semiconductor (CMOS) technology and produce low-cost devices. In this regard, efforts have been made to develop low-loss Si based material for integrated circuits working in the mid-infrared range while bypassing Si strong absorption above 7 $\upmu\text{m}$\cite{chandler-horowitz_high-accuracy_2005}. Germanium (Ge) thus appears to be a very promising candidate since it shows a transparency window of up to 15 $\upmu\text{m}$ and is compatible with monolithic integration on Si and CMOS technology\cite{marris-morini_germanium-based_2018,hu_silicon_2017}. SiGe photonics circuits operating in the mid-infrared have thus been developed, and many devices have been reported \cite{sinobad_mid-infrared_2018,li_ge--si_2019}. Ge-rich SiGe waveguides obtained by gradually increasing the Ge concentration in the waveguide have also been proposed and developed. This gradual increase helps to smoothly adapt the lattice structure from Si to Ge and to benefit from the full transparency of Ge up to 15 $\upmu\text{m}$ wavelength. As an example, two-octave supercontinuum generation from 3 to 13 $\upmu\text{m}$ was obtained showing promising performances\cite{montesinos-ballester_-chip_2020}. 

This work presents a chirped Bragg grating filter implemented within a graded-index SiGe photonic circuit working at 8 $\upmu\text{m}$ wavelength designed to compensate for the chirp of a QCL frequency comb source. The chirped Bragg grating filters have been first designed, fabricated and characterized. Then they have been coupled to a QCL frequency comb, and on-chip pulse generation has been demonstrated for the first time at 8 $\upmu\text{m}$ wavelength.
 
\section{Chirped Bragg grating filter design}
\label{design}
Pulse compression requires a chirped filter capable of compensating for the chirp of the source — in this case, a Fabry–Perot QCL frequency comb. 

Chirped Bragg grating filters are promising candidates for this purpose\cite{tan_monolithic_2010,choi_high_2021}. They are based on Bragg gratings with a non-uniform period such that different wavelengths are reflected at different positions along the grating. As a result, different wavelengths experience different time delays, enabling a wavelength-dependent group delay. Moreover, the overall group delay dispersion (GDD) can be tailored by adjusting the chirped Bragg grating length, allowing precise dispersion control while maintaining compact device dimensions.

\begin{figure}[h]
    \centering
    \includegraphics[scale=0.55]{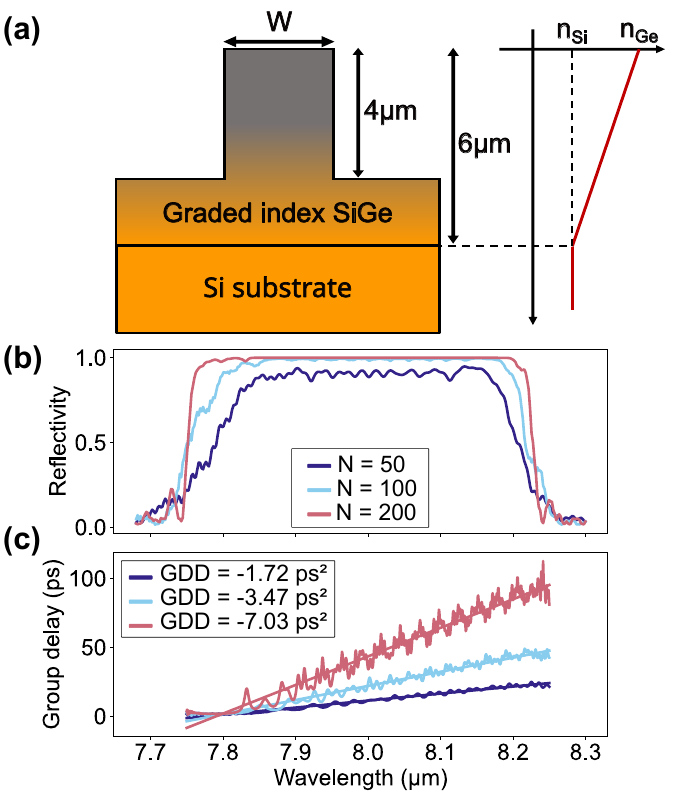}
    \caption{\textbf{(a)} Left: Schematic view of 6 $\upmu\text{m}$ thick graded index SiGe from Si to pure Ge with 4 $\upmu\text{m}$ etching depth waveguide, Right: refractive index profile \textbf{(b)} Simulated reflection and \textbf{(c)} Group delay as a function of the wavelength for N = 50, 100 and 200 showing that the chirped Bragg grating affect wavelengths between 7.8 and 8.2 $\upmu\text{m}$.}
    \label{fig:plateforme+simu}
\end{figure}

A 6 $\upmu\text{m}$ thick graded SiGe photonics platform with linear increase from Si to Ge is used as shown in figure \ref{fig:plateforme+simu}.a. Each Bragg grating period is made of two sections with a width W (see figure \ref{fig:plateforme+simu}.a) equal to 6 $\upmu\text{m}$ or 4 $\upmu\text{m}$. Aiming a bandgap around 8 $\upmu\text{m}$ from $\lambda _i = 7.8$ to $\lambda _f = 8.2$ $\upmu\text{m}$,  the chirped Bragg grating is thus constructed as such: 15 Bragg gratings each with a different period between $\Lambda _i = 1.068$ and $\Lambda _f = 1.128$ $\upmu\text{m}$ are concatenated. Therefore, each Bragg grating reflects a different wavelength range, and by tuning the number of periods of each Bragg grating, the GDD is tuned as well since the different wavelength reflection position will vary. In the rest of the paper the number of periods of each Bragg grating will be referred to as N. 

\begin{figure}[h]
    \centering
    \includegraphics[scale=0.3]{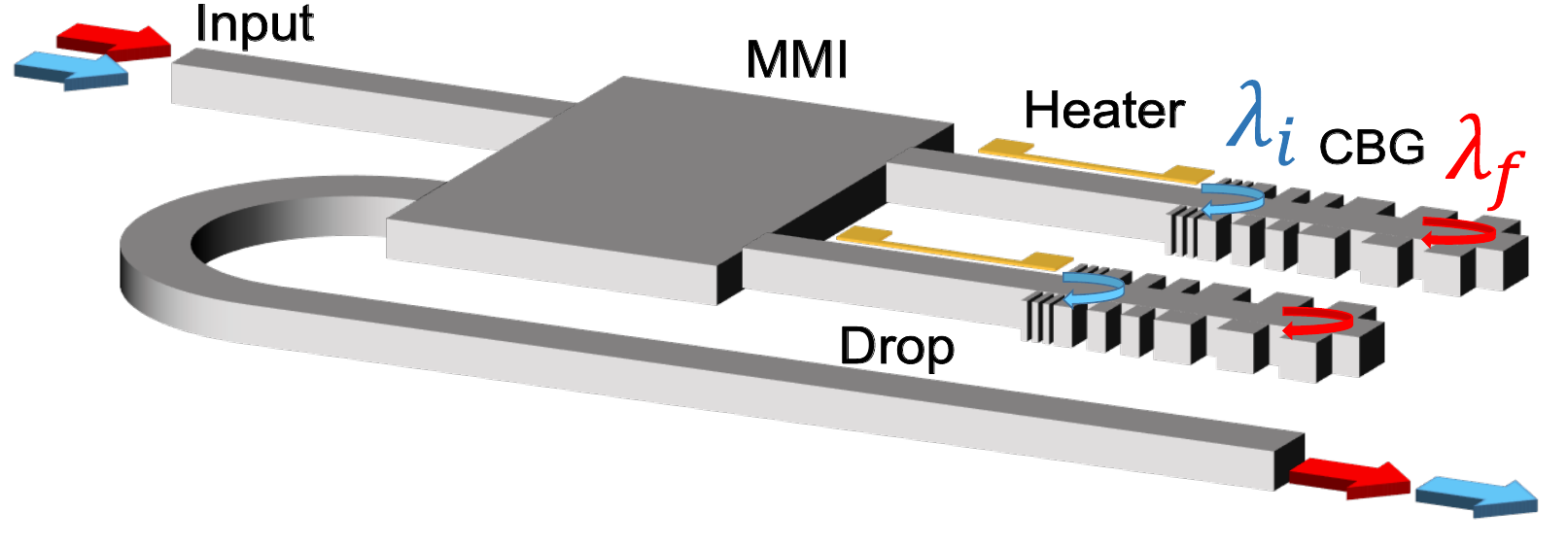}
    \caption{Photonic integrated circuit for pulse compression: Initially chirped light is injected into a symmetric 2 $\times$ 2 MMI where it is evenly split between the two outputs. Each output contains a chirped Bragg grating reflecting and compressing the light. The compressed reflected light re-enters the MMI and all light goes into a drop arm where it is retrieved.}
    \label{fig:PIC}
\end{figure}

Eigen-mode expansion simulations were conducted to calculate the chirped Bragg grating bandgap and group delay as a function of the wavelength. Figure \ref{fig:plateforme+simu}.b shows the reflectivity as a function of the wavelength for N = 50, 100 and 200. We can see that the reflectivity is over 80$\%$ for N = 50 and above. Notably, N = 50 corresponds to a total length of approximately 800 $\upmu$m. The group delay as a function of the wavelength for N = 50, 100 and 200 can be seen figure \ref{fig:plateforme+simu}.c. Linear profiles with GDD of -1.72, -3.47 and -7.03 ps$^2$ for N = 50, 100 and 200, respectively, were extracted. Chirped Bragg gratings can thus be used for pulse compression as they show GDDs of the same amount as the one of the QCL frequency combs\cite{taschler_femtosecond_2021} making this device promising for on-chip pulse compression. However, to do so, the beam reflected by the chirped Bragg grating must be extracted, as it is the one that will undergo compression. This is achieved by using a symmetric 2 $\times$ 2 multimode interferometer (MMI)\cite{vakarin_ultra-wideband_2017} with each output arm containing a chirped Bragg grating\cite{kashyap_laser-trimmed_1993,wang_low_2015} (see figure \ref{fig:PIC}). Light is injected into an input port of the MMI. The power is evenly split between the two output arms with a phase difference of $\frac{\pi}{2}$ because the MMI is symmetric. Reflected light thus re-enters the MMI  allowing all the light to go through a drop port where the reflected beam is collected. The collection of reflected light requires good control of the phase of the beams that interfere at the MMI output. Phase shifters are thus used to correct any phase variation that could occur due to fabrication. They consist of a 10 $\upmu\text{m}$ wide metallic line close to each output waveguide\cite{koompai_long-wave_2023}. This is the scheme that will be used in the following to extract the reflected beam from the chirped Bragg grating filters (see figure \ref{fig:PIC}).  

\section{Results}
\subsection{Fabrication process flow}
\begin{figure}[h]
    \centering
    \includegraphics[scale=0.45]{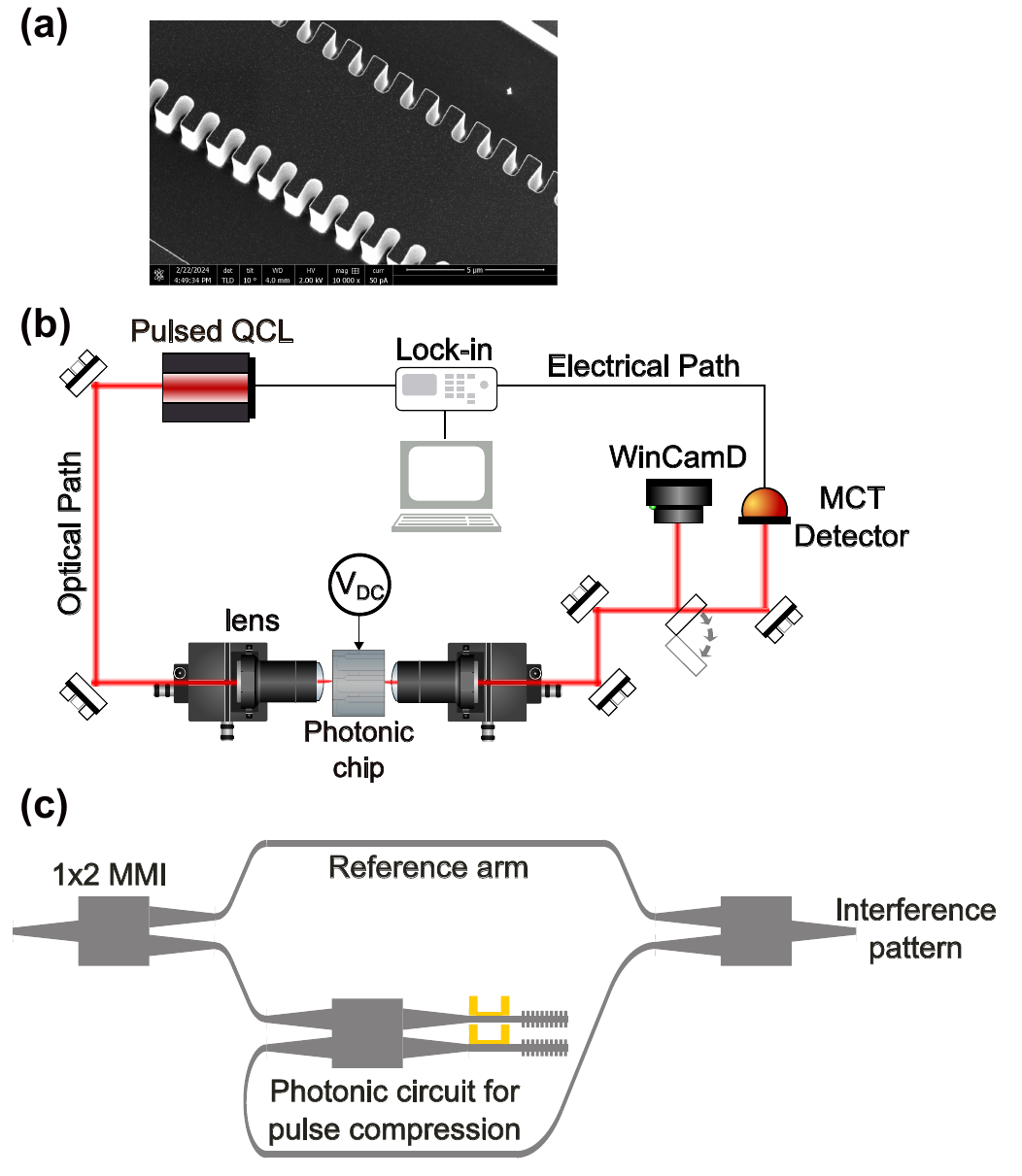}
    \caption{\textbf{(a)} SEM image of a chirped Bragg grating. \textbf{(b)} Characterization set-up: light from a QCL laser is coupled into the waveguides by a ZnSe input lens. The output beam is collected by an other lens and directed into a MCT detector where the transmission spectrum is retrieved. \textbf{(c)} Scheme of the integrated MZI used for group delay measurement: light is evenly split between two arms. One is a reference arm and the other the photonic circuit for pulse compression. The two arms are recombined and an interference pattern with oscillation varying with the wavelength is measured}
    \label{fig:set-up+SEM}
\end{figure}

Fabrication begins with epitaxial growth of the graded SiGe layer by low-energy plasma-enhanced chemical vapor deposition on a silicon substrate. The position of the metallic heaters is first defined by electron-beam lithography. Chemical vapor deposition is then performed to deposit a 10 nm thick titanium layer and a 300 nm thick gold layer and is followed by lift-off. Waveguides are then fabricated by a second electron beam lithography followed by inductively coupled plasma reactive ion etching. A 15 $\upmu\text{m}$ wide coupling section followed by a 200 $\upmu\text{m}$ long taper is used for light injection and collection that are performed by butt coupling. Waveguide facets are diced by optical-grade dicing to ease light injection. Scanning electron microscope (SEM) image of the chirped Bragg grating can be seen in figure \ref{fig:set-up+SEM}.a.

\subsection{Reflection and group delay dispersion measurement}
\begin{figure}[h]
    \centering
    \includegraphics[scale=0.35]{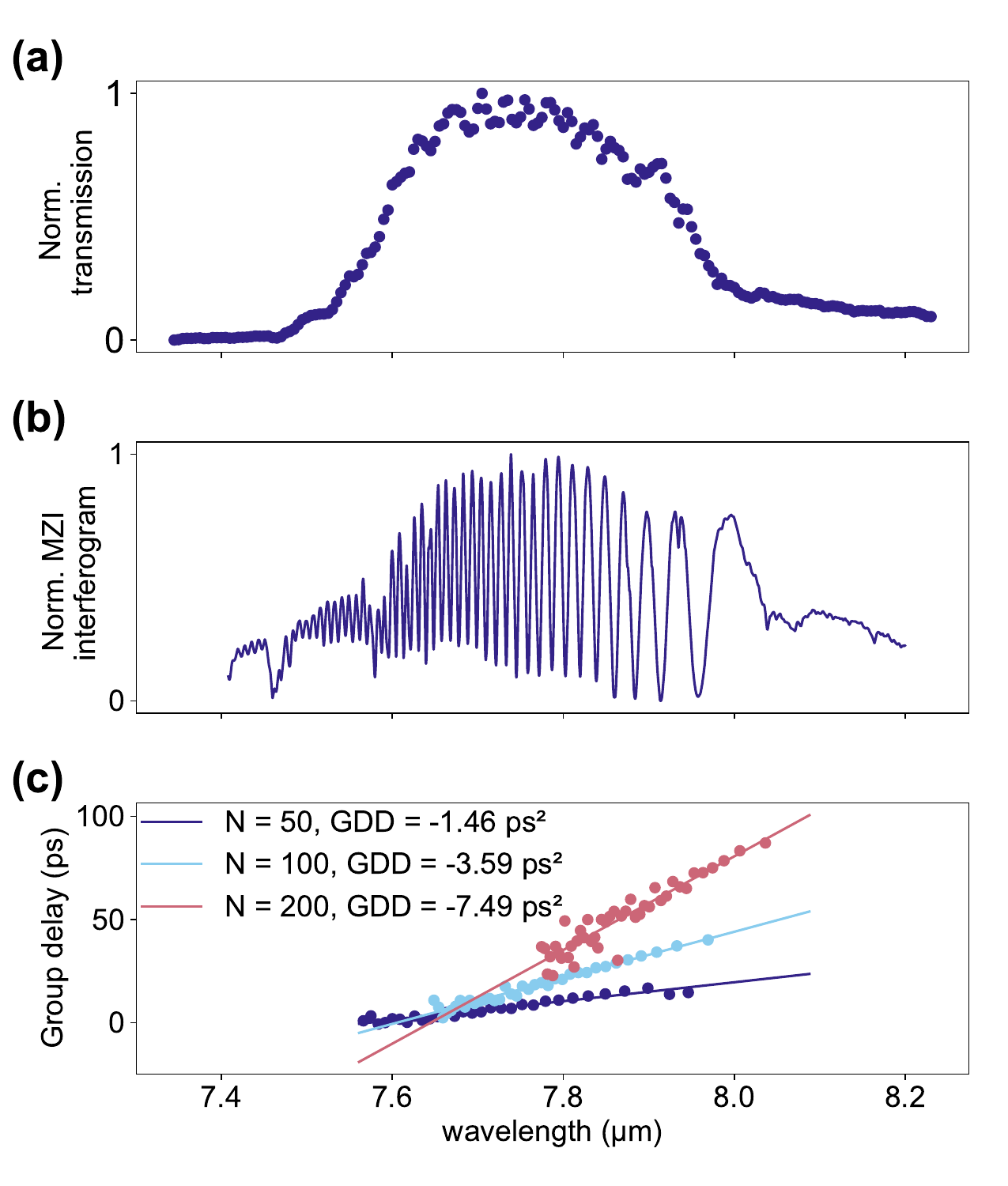}
    \caption{\textbf{(a)} Normalized reflection measurement for N = 50 showing a reflection bandgap between approximately 7.5 and 8 $\mu$m \textbf{(b)} MZI interferogram exhibiting oscillations with an increasing period with the wavelength \textbf{(c)} Group delay extracted from MZI interferograms showing linear group delay dependance with the wavelength and GDD = -1.46, -3.59 and -7.49 ps$^2$ for N = 50, 100 and 200 respectively.}
    \label{fig:charac-results}
\end{figure}

A schematic view of the characterization set-up is given figure \ref{fig:set-up+SEM}.b. Ligth from a tunable QCL is focalized on the waveguide input by an aspheric Zinc-Selenide (ZnSe) lens. An identical lens collects the light from the waveguide output. A bolometer camera is used to observe the optical mode and ensure proper light injection. A Mercury Cadmium Telluride (MCT) detector serves to measure the transmission coming out of the waveguides. The detector is connected to a lock-in amplifier to reduce noise. Reflection measurement from the photonic circuit shown in figure \ref{fig:PIC} for N = 50 is given in the plot figure \ref{fig:charac-results}.a. We observe a 400-nm reflection band around 7.75 $\upmu\text{m}$. This central wavelength can be further tuned by slight adjustments of the Bragg period.

The group delay dispersion is measured using an integrated Mach-Zehnder interferometer (MZI). One arm of the MZI is a reference arm and the other one contains the photonic circuit for pulse compression (see figure \ref{fig:set-up+SEM}.c). In our case, the optical path difference will thus vary with the wavelength resulting in oscillations that also vary with the wavelength in the interference pattern as can be seen figure \ref{fig:charac-results}.b. The group delay as a function of the wavelength can be extracted from this curve using this formula: $\tau _g = (\lambda ^2 +\Delta \lambda n_g \Delta L_0)/(c\Delta \lambda)$, with $\tau _g$ the group delay, $\lambda$ the wavelength, $\Delta \lambda$ the oscillations period, $n_g$ the group index, c the light velocity, and $\Delta L_0$ the path length difference considering uniform reflection for all wavelengths in the chirped Bragg grating. Interestingly, the extracted group delay presents a linear shape as we can see figure \ref{fig:charac-results}.c. Measured GDD of -1.46, -3.59, -7.49 ps$^2$ for N = 50, 100, 200 respectively appears to be in very good agreement with the simulations. We can see that the group delay was not measured in the same intervals for all N values because our laser resolution could not allow all oscillations to be measured. In parallel, the photonic circuit losses between the input and the drop waveguides (figure \ref{fig:PIC}) were evaluated to be between 3.3 dB for N = 200 and 5 dB for N=50. These devices show very good prospects for pulse compression measurements and their combination with QCL frequency comb sources is described in the next section.

\subsection{Pulse compression measurement}
\begin{figure}
    \centering
    \includegraphics[scale=0.4]{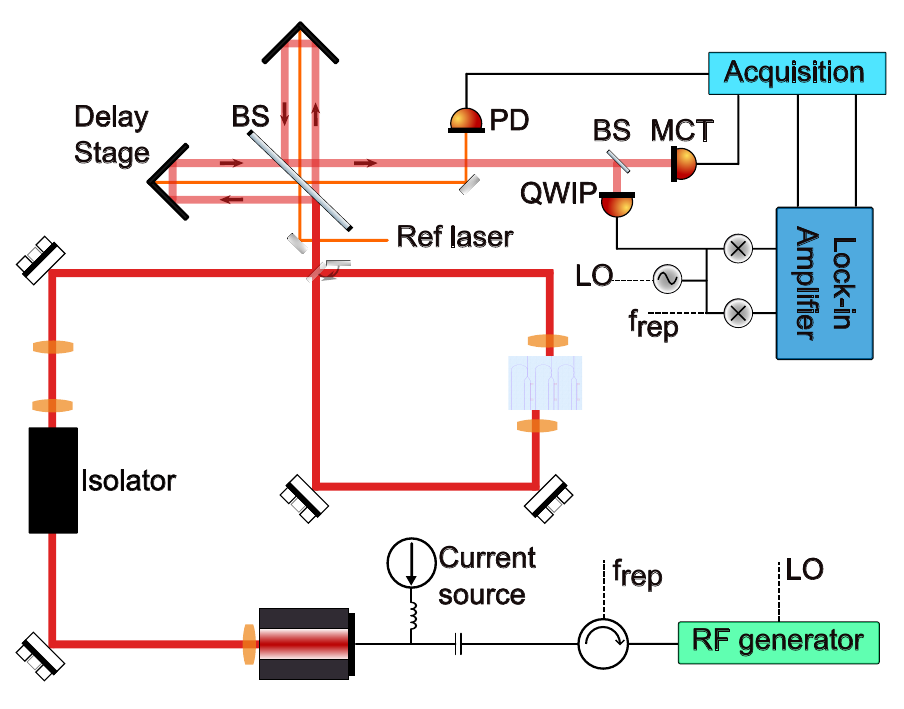}
    \caption{Light from the QCL frequency comb first goes into an isolator to prevent feedback. The beam is then collimated to be coupled into the SiGe photonic circuit and collected out with lenses. The collected beam is then sent into the folded MZI. The interferogram from the MZI is measured by a slow MCT detector to extract the amplitude spectrum and by a fast QWIP to extract the phase. The signal from the QWIP as well as the signal from the QCL are downmixed to be demodulated by a lock-in amplifier. An RF generator is used to lock the QCL to stabilize it and to generate the local oscillator for downmixing. A reference 1.55 $\upmu\text{m}$ laser is used to align the MZI and for data sampling.}
    \label{fig:SWIFTS-set-up}
\end{figure}

Pulse characterization was conducted here through a frequency domain method followed by a reconstruction of the temporal profile. Information about the phase and the spectral amplitudes is necessary to verify that the QCL still behaves as a comb after compression and that the phase was efficiently compensated. Hence, the shifted wave interference Fourier transform spectroscopy (SWIFTS) is employed here\cite{burghoff_evaluating_2015,burghoff_terahertz_2014}. This method is similar to a Fourier transform infrared spectroscopy but coherent lock-in detection with a fast detector is performed to measure the phase information. The full measurement set-up can be seen figure \ref{fig:SWIFTS-set-up}. SWIFTS measurements were conducted for the laser before and after pulse compression i.e. with or without coupling into the photonic circuit for pulse compression. For the compressed pulse measurement, the beam from the QCL comb is coupled into the waveguides using a lens. Another lens is put at the waveguide output to collect the compressed pulse and direct it to a folded MZI. The interferogram from the folded MZI is measured by a fast quantum well infrared photodetector (QWIP). The signal from the QWIP is downmixed with the beat note from the QCL comb ($\textnormal{f}_{\textnormal{rep}}$) using a local oscillator (LO). The laser beatnote is locked with an RF generator to ensure stability for measurements. The local oscillator is produced by the same RF generator with a frequency chosen according to the beatnote to fall into the bandwidth of the lock-in amplifier. The signals can then be demodulated by the lock-in amplifier and phase information is retrieved from the two quadrature signals. The amplitude spectrum is measured in parallel with an MCT detector for its better sensitivity.  

\begin{figure*}
    \centering
    \includegraphics[width=\linewidth]{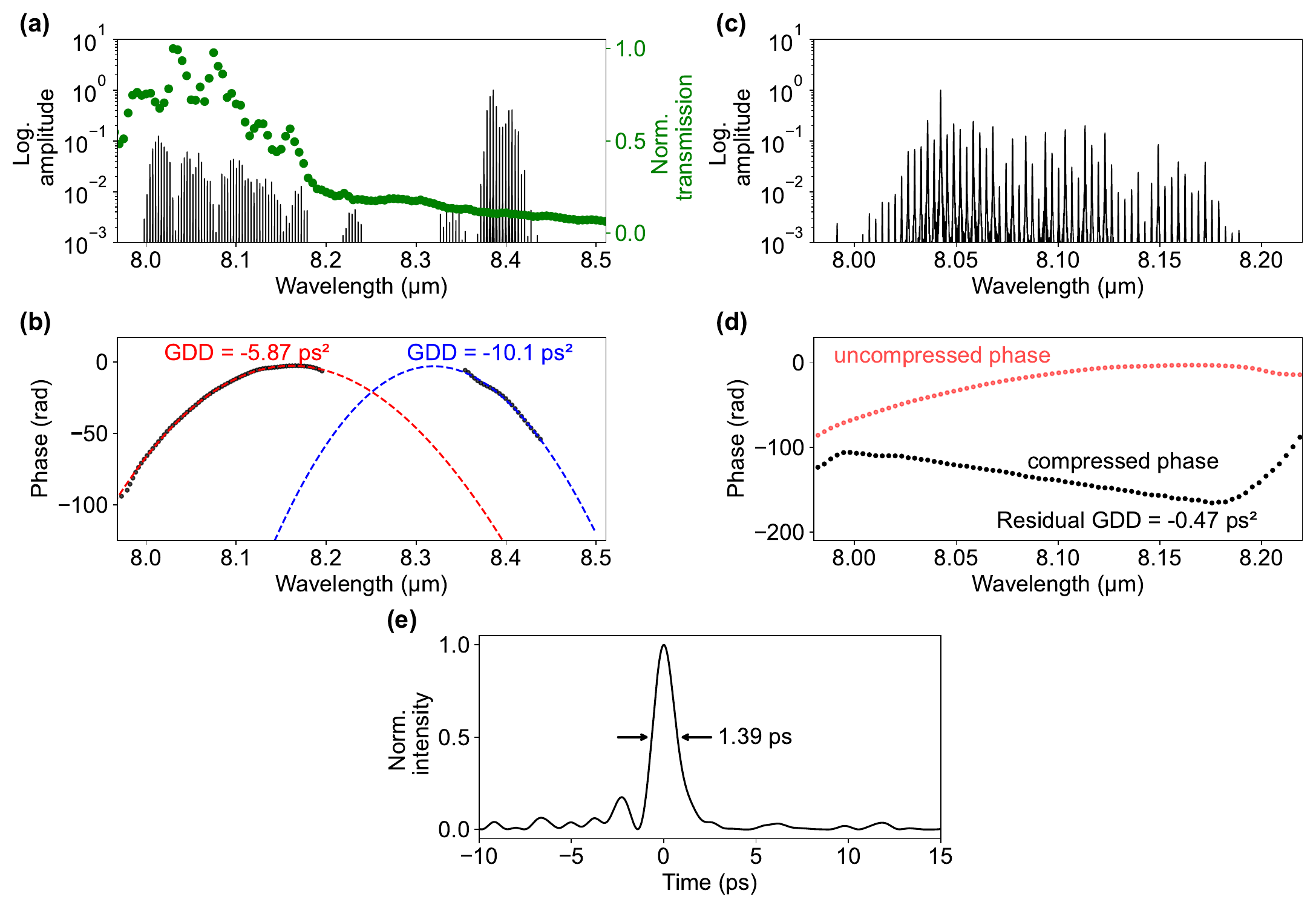}
    \caption{\textbf{(a)} QCL frequency comb amplitude spectrum exhibiting two spectral lobes from 8 to 8.2 $\mu$m and from 8.37 to 8.42 $\upmu\text{m}$. The amplitude spectrum is compared to the SiGe photonics circuit transmission (in green) which matches the left lobe. \textbf{(b)} Phase from the QCL comb showing quadratic dependance with wavelength for both lobes. The shorter wavelength lobe exhibits a GDD = -5.87ps$^2$ and the longer wavelength lobe a GDD = -10.1ps$^2$. \textbf{(c)} Amplitude spectrum after pulse compression with an emission band between 8 and 8.2 $\upmu\text{m}$. \textbf{(d)} Phase after pulse compression (black) illustrating a linear dependance with the wavelength between 8 and 8.18 $\upmu\text{m}$ compared to phase before pulse compression (red). A residual GDD = -0.47 ps$^2$ was measured. \textbf{(e)} Extracted temporal intensity profile exhibiting a 1.39 ps wide pulse}
    \label{fig:SWIFTS-measurements}
\end{figure*}

The QCL comb spectral amplitudes and phase before pulse compression are shown in figure \ref{fig:SWIFTS-measurements}.a and b respectively. Two main lobes can be observed: one between 8 and 8.18 $\upmu\text{m}$ and the other between 8.37 and 8.42 $\upmu\text{m}$. The two lobes have a quadratic phase profile as anticipated \cite{taschler_femtosecond_2021}. The shorter wavelength lobe has a GDD = -5.84 ps$^2$ and the longer wavelength lobe has a GDD = -10.1 ps$^2$.

A new photonic circuit was designed to match the QCL comb properties properly. A chirped Bragg grating with a bandgap corresponding to the shorter wavelength lobe (see figure \ref{fig:SWIFTS-measurements}.a) was used because it is spectrally wider and thus we can expect shorter pulses. The chirped Bragg grating was designed to have a GDD = 5.4 ps$^2$. It was engineered similarly to the chirped Bragg gratings presented in section \ref{design}, but here the chirped Bragg grating is flipped to invert the GDD sign and N = 145 to obtain such a GDD value. SWIFTS measurements show that the higher wavelength spectral lobe was filtered by the chirped Bragg grating and only the shorter wavelength lobe remains as seen in figure \ref{fig:SWIFTS-measurements}.c. In addition, figure \ref{fig:SWIFTS-measurements}.d shows the phase profile after pulse compression exhibiting that the quadratic phase profile was compensated to a linear phase profile where the amplitude spectrum is above the noise level which can be interpreted as a time delay. By detrending the phase curve and fitting a second order polynom, we evaluated that the initial chirp was reduced to a residual GDD of -0.47 ps$^2$. The temporal intensity profile is extracted from the phase and the spectral amplitudes and is shown figure \ref{fig:SWIFTS-measurements}.e. A pulse with a 1.37 ps full width at half maximum is obtained, showing efficient pulse compression. The intermodal coherence after pulse compression can be found in the Supplementary materials. 

\newpage

\section{Conclusion}
In this work, we successfully demonstrated temporal compression of a QCL frequency comb using an integrated photonic circuit. By employing chirped Bragg gratings, we compressed an initially quasi-continuous temporal intensity profile into pulses as short as 1.39 ps. This result marks a step toward fully integrated photonic circuits for supercontinuum generation driven by temporally compressed QCL combs. 

\section*{Supplementary materials}
Please refer to the Supplementary materials for information about the coherence of the SWIFTS measurements.

\begin{acknowledgments}
This work was partly supported by the French RENATECH network. This project has received funding from the European Union's Horizon Europe research and innovation program (101128598 - UNISON) and from the European Research Council (101097569 - Electrophot). Views and opinions expressed are however those of the authors only and do not necessarily reflect those of the European Union or the European Research Council. Neither the European Union nor the granting authority can be held responsible for them.
\end{acknowledgments}

\section*{Conflict of Interest Statement}
The authors have no conflicts to disclose.

\section*{Data Availability Statement}
The data that support the findings of this study are openly available at Zenodo repository.

\section*{Author Contributions}
\textbf{Annabelle Bricout:} Conceptualization (equal); Methodology (equal); Writing/Original Draft Preparation (equal); Software (equal).
\textbf{Mathieu Bertrand:} Conceptualization (equal); Methodology (equal); Software (equal).
\textbf{Philipp Täschler:} Conceptualization (equal); Methodology (equal); Software (equal).
\textbf{Barbara Schneider:} Conceptualization (equal); Methodology (equal).
\textbf{Victor Turpaud:} Conceptualization (equal); Software (equal).
\textbf{Stefano Calcaterra:} Conceptualization (equal); Resources (equal)..
\textbf{Davide Impelluso:} Conceptualization (equal); Resources (equal).
\textbf{Marco Faverzani:} Conceptualization (equal); Resources (equal).
\textbf{David Bouville:} Resources (equal).
\textbf{Jean-René Coudevylle:} Resources (equal).
\textbf{Samson Edmond:} Resources (equal).
\textbf{Etienne Herth:} Resources (equal).
\textbf{Carlos Alonso-Ramos:} Conceptualization (equal); Writing/Review \& Editing (equal).
\textbf{Laurent Vivien:} Conceptualization (equal); Writing/Review \& Editing (equal).
\textbf{Jacopo Frigerio:} Writing/Review \& Editing (equal); Resources (equal).
\textbf{Giovanni Isella:} Writing/Review \& Editing (equal); Resources (equal).
\textbf{Jérôme Faist:} Conceptualization (equal); Writing/Review \& Editing (equal); Methodology (equal).
\textbf{Delphine Marris-Morini:} Conceptualization (equal); Writing/Review \& Editing (equal); Methodology (equal); Funding Acquisition (lead).


\bibliography{bibliography}

\end{document}